\documentclass[prl,preprint,superscriptaddress,aps,floatfix]{revtex4}
\usepackage{graphicx}
\usepackage{color}
\usepackage{bm}
\usepackage{subfigure}

\begin{document}

\title{The Single-Channel Regime of Transport through Random Media}

\author{A.~Pe\~{n}a}
\affiliation{Department of Physics and Astronomy, University of Texas at San Antonio, San Antonio, TX 78249, USA}

\author{A.~Girschik}
\affiliation{Institute for Theoretical Physics, Vienna University of Technology, A--1040 Vienna, Austria, EU}

\author{F.~Libisch}
\affiliation{Department of Mechanical and Aerospace Engineering, Princeton University, Princeton, NJ 08544, USA}

\author{S. Rotter}
\affiliation{Institute for Theoretical Physics, Vienna University of Technology, A--1040 Vienna, Austria, EU}

\author{A.~A.~Chabanov}
\affiliation{Department of Physics and Astronomy, University of Texas at San Antonio, San Antonio, TX 78249, USA}

\begin{abstract}
The propagation of light through samples with random inhomogeneities can be described by way of transmission eigenchannels, which connect incoming and outgoing external propagating modes. Although the detailed structure of a disordered sample can generally not be fully specified, these transmission eigenchannels can nonetheless be successfully controlled and utilized for focusing and imaging light through random media. Here we demonstrate that in deeply localized quasi-1D systems, the single dominant transmission eigenchannel is formed by an individual Anderson localized mode or by a ``necklace state''. In this single-channel regime, the disordered sample can be treated as an effective 1D system with a renormalized localization length, coupled through all the external modes to its surroundings. Using statistical criteria of the single-channel regime and pulsed excitations of the disordered samples allows us to identify long-lived localized modes and short-lived necklace states at long and short time delays, respectively.
\end{abstract}

\maketitle

Because disorder is widespread in natural and artificial materials, transport through random media is of broad interest. This subject encompasses the propagation of classical electromagnetic and mechanical waves as well as the transport of quantum mechanical matter waves in a wide variety of systems \cite{Sebbahbook,Wiersma,Mesobook,Billy,Roati,Kondov}. Recent advances in imaging and focusing of light through strongly scattering media have attracted particular attention as they hold great promise for a host of practical applications (for a review, see \cite{MoskReview} and references therein). At the heart of all studies dealing with wave propagation through random media, lies essentially a scattering problem which can be approached by methods of mesoscopic transport theory (see, e.g., \cite{Beenakker}). In this theoretical framework, an incoming wave is decomposed into transverse free-propagating modes, associated with the quantized directions at which the incoming wave is incident onto the disordered sample (similarly for the outgoing wave). The number of transverse modes $N$ is directly proportional to the area $A$ of the sample cross-section, $N\propto A/\lambda^2$, where $\lambda$ is the wavelength. These modes define the basis for the $N\times N$ transmission matrix $t$ which provides a full description of wave transmission through the sample. The transmittance of the sample can be characterized by the so-called transmission eigenvalues $\tau_n$, $n=1,\ldots,N$, of the Hermitian matrix $tt^{\dagger}$, each of which lies between 0 (no transmission) and 1 (perfect transmission). The statistics of these transmission eigenvalues and of the associated eigenchannels strongly depends on the specific transport regime the disordered sample is in:  In the diffusive regime, for sample lengths $L$ much less than the localization length $\xi$, a fraction $N'\approx\xi/L$ of the total $N$ transmission eigenchannels, on average, feature a $\tau_n$ close to unity (open channels), whereas the rest of the channels have exponentially small transmission (closed channels) \cite{Dorokhov,Imry}. Beyond the localization length $\xi$, at which $N'\approx 1$, transmission becomes dominated by a single eigenchannel associated with the largest transmission eigenvalue $\tau_1$, which falls off exponentially with $L$ due to Anderson localization \cite{Anderson,Vasiliev}. Note that for a given random sample, the transmission eigenvalues and eigenchannels can be determined only after the entire transmission matrix $t$ is measured. Until very recently, such a measurement seemed a formidable task for optical samples given the tremendous size of their transmission matrix (with $N\gg 1$). This challenge has been nearly overcome \cite{Popoff} when a remarkably large portion of the optical transmission matrix was recorded and utilized for focusing and imaging light through random media \cite{Mosk,Mosk1,Popoff1,Katz}. Even more recently, the transmission matrix was measured in its entirety for microwaves transmitted through quasi-1D random waveguides \cite{GenackPRL} and computed numerically for optical waves propagated through 2D localized  slabs \cite{Choi12}. In quasi-1D diffusive samples the transmitted field was found to be a random admixture of many transmission eigenchannels, whereas in localized systems transmission was generally dominated by a single eigenchannel.

While the transmission eigenchannels are convenient for describing transport through random media, one would like to know how these channels are formed in a given random sample. This information is contained, in principle, in the relation between the transmission eigenvalues $\tau_n$ and the transmission matrix $t=u \cdot\sqrt{\mathrm{diag}(\tau_n) }\cdot v$ \cite{Mello}. The unitary $N\times N$ matrices $u$ and $v$ define here the transmission eigenchannels between linear combinations of the incoming and outgoing {\it external} modes of the system. This definition, however, implicitly includes the details of the contacts to the asymptotic regions. An alternative way to determine the transmission eigenchannels, which is both less arbitrary and more insightful, is to refer to the {\it internal} modes in a random medium. These internal modes are characterized by the Thouless number \cite{Thouless,Gang4}, $\delta\equiv\delta\nu/\Delta\nu$, where $\delta\nu$ is the typical spectral width of the modes in the medium and $\Delta\nu$ is the average spacing between neighboring modes. The Thouless number $\delta$ is thus a measure of the degree of spectral mode overlap, with the diffusive and the localized regime being hallmarked by overlapping modes ($\delta>1$) and well-resolved modes ($\delta<1$), respectively. In the localized regime, where at a given frequency just a single or a few internal modes participate in the transport process, it has recently been shown experimentally that the transmission spectrum of a disordered medium can be fully characterized by such internal modes \cite{GenackNature}.

The fact, that in the localize regime the dominance of only a single or a few internal modes concurs with the dominance of a single transmission eigenchannel, suggests that some deeper relation between these complementary pictures of transport exists. Indeed, recent numerical work showed that internal modes and transmission eigenchannels displayed strong correlations with each other \cite{Choi12}. Here, we make an important step further and uncover a {\it direct} link between modes and channels. To reveal this link unambiguously, we probe into the deeply localized limit where just a {\it single} transmission eigenchannel dominates. In this experimentally and numerically challenging limit, we demonstrate that the dominant transmission eigenchannel is formed by an {\it individual} localized mode or by a {\it unique} superposition of localized modes giving rise to a so-called ``necklace state’’ \cite{Pendry,Pendry1D}. To establish this connection between the eigenchannel and mode pictures of transport, we investigate random quasi-1D samples with a length $L$ that is significantly larger than the localization length $\xi$. The numerical and experimental results which we obtain in this deeply localized limit allow us to identify both the characteristic signatures of the single-channel regime as well as its connection to the internal modes of the disordered sample.

\noindent{\bf Results}

\noindent{\bf Crossover to the single-channel regime.} In a scattering experiment with light or microwaves, a monochromatic wave is typically incident in a given direction and the transmitted wave can be detected in any direction, so that the individual elements of the transmission matrix, $t_{ab}$, (between incident and outgoing modes $a$ and $b$, respectively) can be measured directly for a given sample. From these matrix elements $t_{ab}$, the following three types of transmittances can be computed,
\begin{equation}
T_{ab}=|t_{ab}|^2\,, \,\,\,\, T_a=\sum_{b=1}^{N} |t_{ab}|^2\,, \,\,\,\, T=\sum_{a,b=1}^{N} |t_{ab}|^2\,.
\end{equation}
The transmittance $T_{ab}$, for a given incident mode $a$ and outgoing mode $b$, gives rise to a speckle pattern in the transmitted intensity. The transmittance $T_a$ is the total transmission (for the incident mode $a$), and thus represents the apparent brightness of the speckle pattern. The transmittance $T$ is the classical analogue of the dimensionless conductance $g$ \cite{Landauer,Fisher}.

In the single-channel regime, these key transmittances are all essentially determined by the largest transmission eigenvalue $\tau_1$. Using again the above decomposition $t = \sum u \cdot \mathrm{diag}(\sqrt{\tau_n})\cdot v$, we can write, $T_{ab}=|u_{a1}|^2\tau_1|v_{1b}|^2$, $T_a=|u_{a1}|^2\tau_1$, and $T=\tau_1$. From the expression for $T_{ab}$, one can see that in a given random sample (with a given $\tau_1$) the speckle pattern of transmitted intensity is literally frozen. Altering the incident wave has little or no effect on the positions and relative intensities of speckles, affecting only the apparent brightness of the speckle pattern \cite{GenackNature}. This is in striking contrast to the diffusive regime, in which a minor change in the incident wave leads to a very different speckle pattern \cite{Freund}. To express this difference in statistical terms, we switch to the normalized transmission coefficients, $s_{ab}=T_{ab}/\langle T_{ab}\rangle$, $s_{a}=T_{a}/\langle T_{a}\rangle$, and $s=T/\langle T\rangle$, where $\langle\ldots\rangle$ represents the average over an ensemble of random sample configurations. From the expressions for $T_{ab}$, $T_a$ and $T$ in the single-channel regime, and from negative exponential statistics of $|u_{a1}|^2$ and $|v_{1b}|^2$ \cite{Beenakker}, namely $\langle |u_{a1}|^{2n}\rangle/\langle |u_{a1}|^2\rangle^n=\langle |v_{1b}|^{2n}\rangle/\langle |v_{1b}|^2\rangle^n =n!$ \cite{Goodman}, one can derive the relations between the statistical moments of the normalized transmittances, $\langle s_{ab}^n\rangle=n!\langle s_{a}^n\rangle=(n!)^2\langle s^n\rangle$. As a result, the fluctuations of the normalized intensity are four times larger than those of the normalized transmittance, $\langle s_{ab}^2\rangle/\langle s^2\rangle=4$. In the following, we use exactly this statistical ratio, $R=\langle s_{ab}^2\rangle/\langle s^2\rangle$, to chart the crossover from the diffusive to the single-channel regime \cite{note}.

In our experiment, we employ a microwave setup (see Methods) to measure microwave fields transmitted through quasi-1D random samples of alumina (Al$_2$O$_3$) spheres contained in a long copper tube. The number $N$ of transmission channels associated with the area $A$ of the tube cross-section changes from 24 to 32 over the measured frequency range. The measurements are carried out in random ensembles of 15,000 sample realizations at three different lengths and two alumina filling fractions (Samples $A-D$, see Methods). From the measured field spectra, we compute $\langle s_{ab}^2\rangle$ and $\langle s^2\rangle$, which give us the corresponding values of $R$ and $L/\xi$ (see Methods). The ratio $R$ is plotted versus $L/\xi$ in Fig.~1. Data points of the same color and style are obtained for different frequencies in Sample $A$ (orange circles), $C$ (red squares) and $D$ (brown triangles). Although we consider alumina samples of two filling fractions and three different lengths, all the data points are seen to fall on a single curve. To explain this universality, we utilize exact non-perturbative calculations \cite{Mirlin} for the first two statistical moments of the dimensionless conductance, $\langle g \rangle$ and $\langle g^2 \rangle$, as a function of $L/\xi$, to obtain an analytical expression for $R$ (see Methods), and find the analytical curve for $R$ (black solid line) to agree well with the experimental data. Most importantly, the analytical result clarifies that $R$ monotonically crosses over from $R=2$ in the classical diffusion limit to $R=4$ in the single-channel regime, with increasing $L/\xi$. The perturbative result for $R$, to leading order in $L/\xi$, $R=2+4L/3\xi$, is also plotted in Fig.~1 (black dotted line).  The excessive data variation observed at $L/\xi\geq 2.5$ is caused by a low signal-to-noise ratio, which is due to the increasing effect of absorption.

To further demonstrate the universality of the crossover to the single-channel regime, we numerically study wave scattering through planar disordered waveguides attached on the left and right to clean semi-infinite leads (see Methods). We model the disorder by randomly placing nonabsorbing dielectric scatterers into the middle portion of a waveguide of width $W$ and length $L=5W$. The width $W$ is chosen such that there are $N=15$ channels in the leads at the frequency of interest. We solve the scattering problem for waves incident from the left lead using the advanced Modular Recursive Green's function method \cite{Rotter, Libisch}, which is based on a finite-difference approximation of the scalar Helmholtz equation. To emulate the crossover to the single-channel regime, calculations of the full transmission matrix $t$ are carried out for ensembles of 2,000 random waveguide realizations at eight different system lengths (at a single scattering frequency). After the localization length $\xi=1.52W$ is obtained from fitting the slope of the logarithmic transmittance, $\langle \ln T\rangle\propto -2L/\xi$, the computed values of $R$ are plotted versus $L/\xi$ as line-connected diamonds in Fig.~1. The numerical data agree well with the experimental data and the analytical result for $R$. Note that, in contrast to the microwave measurements, our numerical model is restricted to two dimensions and describes scalar rather than vector waves. Still, the crossover for $R$ which we obtain in this way closely follows the analytical prediction. We also emphasize that our numerical simulations of the dissipation-free model system fully confirm the convergence $R\rightarrow 4$ in the deeply localized limit.

\noindent{\bf Statistics of the single-channel regime.} The fact that in the deeply localized limit transport is mediated by a single transmission eigenchannel has remarkable consequences for the statistical properties of the single-channel regime. The key insight in this respect is that transmission through a single channel can be mapped onto a strictly 1D system where only one transmission channel exists by default. Such a mapping allows us to predict for the single-channel regime the statistical properties which are already known for 1D systems \cite{Pendry1D,Abrikosov,Melnikov}. Consider, e.g., the probability density distributions $P(s_a)$ and $P(s_{ab})$ in the single-channel regime (equation (3) in Methods), which are entirely determined through $P(T)$ --- a distribution which is known analytically in 1D for arbitrary sample length $L$ \cite{Vasiliev,Abrikosov,Melnikov}. To perform this mapping to 1D, it is tempting to choose the effective 1D system such that it has the same system length $L$ and localization length $\xi$ as in our quasi-1D localized systems. However, for increasing system length $L$, the quasi-1D systems first go through a diffusive regime (with an Ohmic reduction of the transmission) before localization sets in \cite{Feilhauer}. By contrast, in true 1D systems such a diffusive regime is entirely absent: only a single channel participates in transport even in samples of vanishing length. Consequently, one would obtain a different value of the average transmission and thus different statistics of transport in 1D as compared to quasi-1D. Although the mapping between these two situations can only be performed in the localized regime, the presence of a diffusive regime in quasi-1D gives rise to a renormalization of the localization length in 1D. The corresponding effective localization length $\xi'$ is chosen such that the transmission in 1D is the same as in quasi-1D, $\xi'=-2L/\langle \ln T \rangle$, for a given $L$, which yields a larger $\xi'$ as compared with the true localization length $\xi$. Note that $\xi'$ is $L-$dependent and approaches $\xi$ for increasing system length $L\to\infty$ (see the inset of Fig.~2, {\bf a}). To explicitly test the above renormalization, we compare both our numerical and experimental results to predictions for the probability density of transmission in 1D from Ref.~\cite{Gopar},
\begin{equation}
P(T)=C\, {\sqrt{{\rm arccosh}(T^{-1/2})}\over T^{3/2}(1-T)^{1/4}}\exp[-{\xi'\over 2L}{\rm arccosh}^2(T^{-1/2})]\,,
\end{equation}
where $C$ is a normalization constant. From this formula we derived an expression for $P(s_{ab})$ and $P(s)$ (see methods, Eq. (4)). Employing here the renormalized localization length $\xi'$ from above, we find excellent agreement between the predictions and our numerical results for the planar system of $L/\xi=5.25$ (see Fig.~2, {\bf a}), notably without any free parameters. In a situation where measuring the absolute transmission is a challenge (as in our experimental setup), $\xi'$ can be found by fitting the transmission distributions from equation (4) (e.g., the probability density $P(\ln s_{ab})$) to the data, as in Fig.~2, {\bf b} for the quasi-1D system of $L/\xi=2.52$. We remark that the discrepancy in Fig.~2, {\bf b}, between the predicted and the experimental distributions (see the tails  of the curves), is due to subdominant transmission eigenchannels and/or a low signal-to-noise ratio. Our successful mapping onto an effective 1D system allows us to reinterpret a disordered system in the single-channel regime as a 1D system diffusively coupled to its surroundings. Since the coupling extends to all the external free-propagating modes, the localization length $\xi'$ becomes renormalized, $\xi' > \xi$.

\noindent{\bf Relationship between the eigenchannels and internal modes.} While the above results provide unique {\it statistical} signatures for the single-channel regime, we now address the question how these transmission channels are formed on a {\it system-specific} level. To answer this question, we turn to our numerical simulations for transmission through a disordered planar waveguide of length $L=5W$ (corresponding to $L=3.29\xi$). A typical spectrum of the transmittance $T$, obtained in a frequency interval of this localized regime, features sharp peaks (see Fig.~3, {\bf a}) resulting from resonant transmission through localized photonic states in the medium. At all frequencies within this range, the transmittance is dominated by the first (largest) transmission eigenvalue $\tau_1$. We verify this explicitly by diagonalizing the Hermitian matrix $tt^\dagger$ and finding that the second largest transmission eigenvalue $\tau_{2}$ fulfills $\tau_2<0.01$ throughout the entire frequency interval of Fig.~3, {\bf a}. In the next step, we examine how these single-channel transmission resonances correspond to specific internal modes inside the disordered region. We consider the same discretized version of the Helmholtz equation as for the scattering problem, but solve for internal modes of the disordered region. We impose constant-flux outgoing boundary conditions \cite{Tureci} at those positions where, in the scattering problem, the external leads are attached. These outgoing boundary conditions make the corresponding eigenvalue problem non-Hermitian, yielding complex eigenfrequencies ${\nu}_m$ that correspond to those of quasi-bound resonances. Following Breit-Wigner theory, the real part of these eigenfrequencies corresponds to the resonance positions $k_m$, $k_m=\mbox{Re}\,{\nu}_m $, while the imaginary part is associated with leaking radiation out of the system, i.e., with the resonance widths, $\Gamma_m=-2\mbox{Im}\,{\nu}_m$. To verify this correspondence explicitly, we reproduce the transmittance $T(\nu)$ in the entire frequency interval of Fig.~3, {\bf a} with a sum of Lorentzian curves,
\begin{equation}
T(\nu)=\left|\sum_{m=1}^N C_m \frac{\Gamma_m/2} {\Gamma_m/2+i(\nu-k_m)}\right|^2\,,
\end{equation}
where $k_m$ and $\Gamma_m$ are determined from the complex eigenfrequencies $\nu_m$, and only the complex amplitudes $C_m$ are fit parameters. The excellent agreement between $T(\nu)$ and the fit (red and black curve, respectively, in Fig.~3, {\bf a}) demonstrates that the position and width of each resonant peak in $T(\nu)$ are, indeed, directly determined by the real and imaginary parts of a single individual mode or a well-defined superposition of very few internal modes. Since, in the single-channel regime, each transmission peak is associated with a transmission eigenchannel, we arrive at the conclusion that each transmission eigenchannel is in turn supported by a unique localized mode or a combination of a few spectrally overlapping modes (marked $b$ and $c$ in Fig.~3, {\bf a}, respectively).

The above correspondence has far-reaching consequences, e.g., in terms of the wave functions (i.e., the spatial field profiles) associated with internal modes and transmission eigenchannels. In particular, the scattering wave functions associated with their respective transmission eigenchannels should correspond to the same linear superposition of internal modes as determined above through the transmittance $T(\nu)$. To test this explicitly, we created a corresponding superposition of internal mode wave functions using the expansion coefficients $C_m$ of equation (3) extracted from the fit to $T(\nu)$ and compared this superposition state to the wave function of the transmission eigenchannel (see plots in Fig.~3, {\bf b, c}). The eigenchannel wave functions are obtained by projecting the numerically calculated Green’s function inside the disordered region onto the external mode configuration of the transmission eigenchannel determined already earlier (see \cite{Rotter,Libisch}). In the case where a transmission eigenchannel is supported by a single internal mode ($b$ in Fig.~3, {\bf a}), we find that the wave function of the mode very well matches that of the scattering state associated with $\tau_1$ at the resonance frequency (see Fig.~3, {\bf b}). Small deviations between the wave functions, visible primarily close to the left boundary of the disordered sample, can be attributed to the fact that the modes feature purely outgoing boundary conditions, whereas the transmission eigenchannels additionally contain the incoming flux from the left lead. The insensitivity of the wave functions deep inside the disordered region with respect to such a change in the boundary condition is, in turn, a hallmark of Anderson localization \cite{Thouless}. In the case of two spectrally overlapping modes ($c$ in Fig.~3, {\bf a}), we compare the scattering state associated with $\tau_1$ to the two-mode superposition wave function determined by the expansion coefficients $C_m$ which were fitted to $T(\nu)$ as outlined in the previous paragraph. We again find very good agreement between the scattering state and this two-mode superposition (see Fig.~3, {\bf c}). Note that the two spectrally overlapping modes feature a high degree of spatial correlation with each other due to mode hybridization \cite{Bliokh,Sebbah} (see Fig.~3, {\bf c}), allowing us to identify these overlapping modes as a necklace state \cite{Pendry,Pendry1D}. We thus conclude that multiple spectrally-overlapping modes form a single transmission eigenchannel by merging into a mode necklace.

\noindent{\bf Dynamics of the single-channel regime.} We finally consider the dynamics of the single-channel regime following a pulsed excitation of a random sample. Since in the single-channel regime the transmission eigenchannel is formed by a localized mode or a necklace state, the eigenchannel bandwidth is equal to the resonance width. Consequently, a pulse of incident radiation with a bandwidth less than the resonance separation will transmit only through a single eigenchannel. By contrast, a pulse of a bandwidth which envelops more than one resonance will transmit through multiple eigenchannels, resulting in modal dispersion and non-exponential decay of the pulsed transmission \cite{Hu04}. To chart the crossover to the single-channel regime in the time domain, we again employ the statistical ratio $R$, evaluated now at different time delays $t$. To accomplish this, the temporal response $t_{ab}(t)$ to a Gaussian pulse is obtained by taking the Fourier transform of the field spectrum multiplied by a Gaussian envelope of width $\sigma$. The computed transmitted intensity, $T_{ab}(t)\equiv|t_{ab}|^2$, then yields $\langle T_{ab}(t)\rangle$ and $R(t)$, using the same methods as for the steady state. Consider first $R(t)$ for the case when the pulse bandwidth $\sigma$ lies between the average resonance width $\delta\nu$ and the average resonance separation $\Delta\nu=\delta\nu/\delta $, as realized, e.g., for a quasi-1D system of $L/\xi=2.9$ (Sample $D$), and  $\sigma=5\delta\nu=0.5\Delta\nu$ (see red line in Fig.~4, {\bf a}). As in the steady state, the single-channel regime occurs in the time domain, when $R=4$. Comparing with the corresponding average transmission $\langle T_{ab}(t)\rangle$ (red line in Fig.~4, {\bf b}), we observe that  the single-channel regime sets in following the arrival of the intensity peak, in agreement with the expectation that the incident pulse with $\sigma<\Delta\nu$ is typically transmitted through a single channel. By contrast, in our numerical simulations for a planar system of $L/\xi=3.29$ and larger bandwidth $\sigma>\Delta\nu$, the single-channel regime ($R=4$) is approached at significantly longer time delays (see the green and brown lines in Fig.~4, {\bf c} which feature a different pulse bandwidth $\sigma$ and therefore a different number of excited modes/resonances). From these observations we conclude that the crossover to the single-channel regime occurs at long time delays when all but one localized mode within the pulse bandwidth have leaked out of the sample.  This is further confirmed by microwave measurements in a diffusive quasi-1D system of $L/\xi=0.4$ (Sample $B$), for $\sigma=1.8\delta\nu=3.9\Delta\nu$ (see the blue lines in Fig.~4, {\bf a} and {\bf b}). As seen in Fig.~4, {\bf a}, $R(t)$ at first increases with time delay monotonically from a steady-state value of 2.8 up to 4, and then stays at 4 for longer time delays. This crossover indicates both that in the diffusive system several transmission eigenchannels are open at short time delays and that there exist long-lived, so-called pre-localized modes \cite{Apalkov} supporting the single-channel regime at long time delays.

An intriguing question to ask at this point is whether we can see any signatures of necklace states in the temporal response of our disordered systems. Note, in this context, that necklace states typically correspond to higher and faster transmission \cite{Choi12} as compared to long-lived localized modes, and therefore they are expected to dominate transmission at shorter time delays. Furthermore, if the incident pulse excites multiple resonances, i.e. $\sigma>\Delta\nu$, yet $\sigma$ is smaller than the average spacing between neighboring necklace-state resonances, only one resonance within the pulse bandwidth is typically realized as a necklace state. (A typical spacing of about $5\Delta\nu$ between neighboring necklace-state resonances of the planar disordered waveguide can be extracted from the data in Fig.~2, {\bf a}.) This is confirmed by the results shown in Fig.~4, {\bf c} where we see that for $\sigma=2.3\Delta\nu$, the ratio $R(t)$ (see the green line) starts out with $R\approx 4$ at the arrival time of the peak intensity, and decreases at later times. This value of $R\approx 4$ at short times suggests that transmission is dominated here by a single transmission eigenchannel formed by a necklace state. As, however, these necklace states are rapidly decaying through the sample boundary, transmission of the incident pulse becomes increasingly dominated by long-lived localized modes supporting multiple eigenchannels (as indicated by $R(t) < 4$ at intermediate  time delays). Finally, in the limit of still longer time delays, we return to the single-channel regime where the most localized mode dominates (as indicated by the final increase of $R(t)$ towards $R=4$). To corroborate these arguments, we investigate the same system with a broader Gaussian pulse of width $\sigma=9.2\Delta\nu$ (see the brown line in Fig.~4, {\bf c}). Since this pulse excites typically more than one necklace-state resonance, the value of $R$ at the peak arrival time drops accordingly (down to 3.5), reflecting the increase in the number of effective transmission eigenchannels.

In summary, our experimental and numerical results disclose an intimate relation between the transmission eigenchannels and spectral modes of disordered systems. We have shown that in deeply localized systems, transmission is governed by a single transmission eigenchannel which is formed by an individual localized mode or by a necklace state. These two types of modes represent slow and fast transmission eigenchannels, which can be probed with a pulsed excitation of a judiciously chosen bandwidth at long and short time delays, respectively. The single-channel regime has unique statistical properties (as, e.g., a frozen speckle pattern), which can be utilized to chart the crossover to single-channel transport in both the steady state and in the time domain. Finally, we have demonstrated that in the single-channel regime a quasi-1D localized system can be mapped onto an effective 1D system with a renormalized localization length, coupled to its surroundings via all available external modes. These results are fundamental to understanding the static and dynamic behavior of waves in random media and can be useful in transmitting energy and information through strongly-scattering complex systems.

\noindent{\bf Methods}

\noindent{\bf Microwave setup.} Microwave transmission measurements were carried out in random mixtures of alumina (Al$_2$O$_3$) spheres of diameter 6.4 mm and refractive index 3.14, contained in a long copper tube of diameter 4.4 cm. Low values of the alumina filling fraction, $f$, were produced by embedding the alumina spheres within Styrofoam shells which are almost transparent for microwaves. We utilized four different alumina samples: $A$ (alumina filling fraction, $f=0.064$, and $L=30.5$ cm), $B$ ($f=0.064$, $L=45.7$ cm), $C$ ($f=0.064$, $L=91.4$ cm) and $D$ ($f=0.125$, $L=45.7$ cm). Microwave transmission spectra were measured within the frequency interval 14.3-16.4 GHz, in which the degree of localization could be tuned with frequency. In the measurement, linearly-polarized microwave radiation was launched and received by conical horns placed 20 cm in front of and behind the sample. Microwave field spectra were taken by using a vector network analyzer for 2 sets of polarizations of the transmitter and receiver antennas, ${\bf e_{\it a}}\perp{\bf e_{\it b}}$ and ${\bf e_{\it a'}}\perp{\bf e_{\it b'}}$, obtained by rotating simultaneously both the antennas by $90^0$. The polarization-selective transmission spectra were used in the data analysis to compute the experimental ratio $R$. Once the field spectra were taken, a new sample realization was created by rotating the sample tube about its axis.

\noindent{\bf Numerical simulations.} In our numerical simulations, we utilized a planar disordered waveguide of width $W$ and length $L$ attached on the left and right to two clean semi-infinite leads. We modeled the disorder by randomly placing nonabsorbing dielectric scatterers of diameter $0.041W$ and refractive index 3.14 into the middle portion of the waveguide at a filling fraction of 0.125, keeping a minimum distance of $0.0205W$ between the scatterers. For the time domain calculations, as well as for the probability distributions of $T$ and $s_{ab}$, we calculated transmission spectra for 100 random disorder configurations at a sample length of $L=5W=3.29\xi$. For each disorder realization the transmission was evaluated in a frequency window with altogether $2397$ frequency points. A small portion of such a spectrum is shown in Fig.~3, {\bf a}.

\noindent{\bf Calculation of $R$}. To obtain the experimental ratio $R=\langle s_{ab}^2\rangle/\langle s^2\rangle$, we computed $\langle s_{ab}^2\rangle$ and $\langle s^2\rangle$ from intensities $T_{ab}$ and $T_{a'b'}$ extracted from the cross-polarized transmission measurements \cite{Abe}. In particular, to compute $\langle s^2\rangle$, we used the relation \cite{Feng,Feng1}, $\langle s^2\rangle=\langle s_{ab}s_{a'b'}\rangle_{a\neq a',b\neq b'}$. To obtain the analytical expression for $R$, we utilized the results for $\langle g \rangle$ and $\langle g^2 \rangle$ of the exact non-perturbative calculations of Ref.~\cite{Mirlin}. To express $\langle s_{ab}^2\rangle$, we used $\langle s_{ab}^2\rangle=2\langle s_{a}^2\rangle$ \cite{Kogan} and $\langle s_{a}^2\rangle=1+{\rm var}s_a=-(\xi/\!\langle g\rangle^{2})\,\partial\langle g\rangle/\partial L$ \cite{Zhang}; for $\langle s^2\rangle$, we used $\langle s^2\rangle=\langle g^2\rangle/\langle g\rangle^2$. The analytical curve for $R$ is plotted versus $L/\xi$ as the black solid line in Fig.~1. In the diffusive regime (for $L\ll\xi$), to leading order in $N\gg1$, $\langle s_{ab}^2\rangle\approx 2+4L/3\xi$ and $\langle s^2\rangle\approx 1+2L^2/15\xi^2$ \cite{Feng,Mello1}, and thus the perturbative result for $R$ is $R\approx 2+4L/3\xi$ (shown by the black dotted line in Fig.~1).

\noindent{\bf Probability distributions of the normalized transmittances in the single-channel regime}. From the relations between the statistical moments of the normalized transmittances, $\langle s_{ab}^n\rangle=n!\langle s_{a}^n\rangle=(n!)^2\langle s^n\rangle$, and using corresponding moment generating functions \cite{Goodman}, we obtained the following relations between their probability density functions,
\begin{eqnarray}
P(s_a)&=&\int_0^\infty {ds\over s}P(s)\exp{(-s_a/s)} \, , \nonumber \\
P(s_{ab})&=&2\int_0^\infty {ds\over s}P(s)K_0(2\sqrt{s_{ab}/s}) \, ,
\label{}
\end{eqnarray}
where $K_0(x)$ is a modified Bessel function of the second kind.

\noindent{\bf Acknowledgments}

\noindent The authors would like to thank M.~Brandstetter, A.~Genack, V.A.~Gopar and R.~Voglauer for very helpful discussions. Financial support by the NSF through project No.~ECCS0926035, the Vienna Science and Technology Fund (WWTF) through project No.~MA09-030 and by the Austrian Science Fund (FWF) through project Nos.~SFB-IR-ON F25-P14, SFB-NextLite F49-P10, No.~I1142-N27 (GePartWave) as well as computational resources by the Vienna Scientific Cluster (VSC) are gratefully acknowledged.
\pagebreak
{\bf FIGURES:}
\begin{figure}
\includegraphics[angle=0, scale=1.0, width=0.7\textwidth]{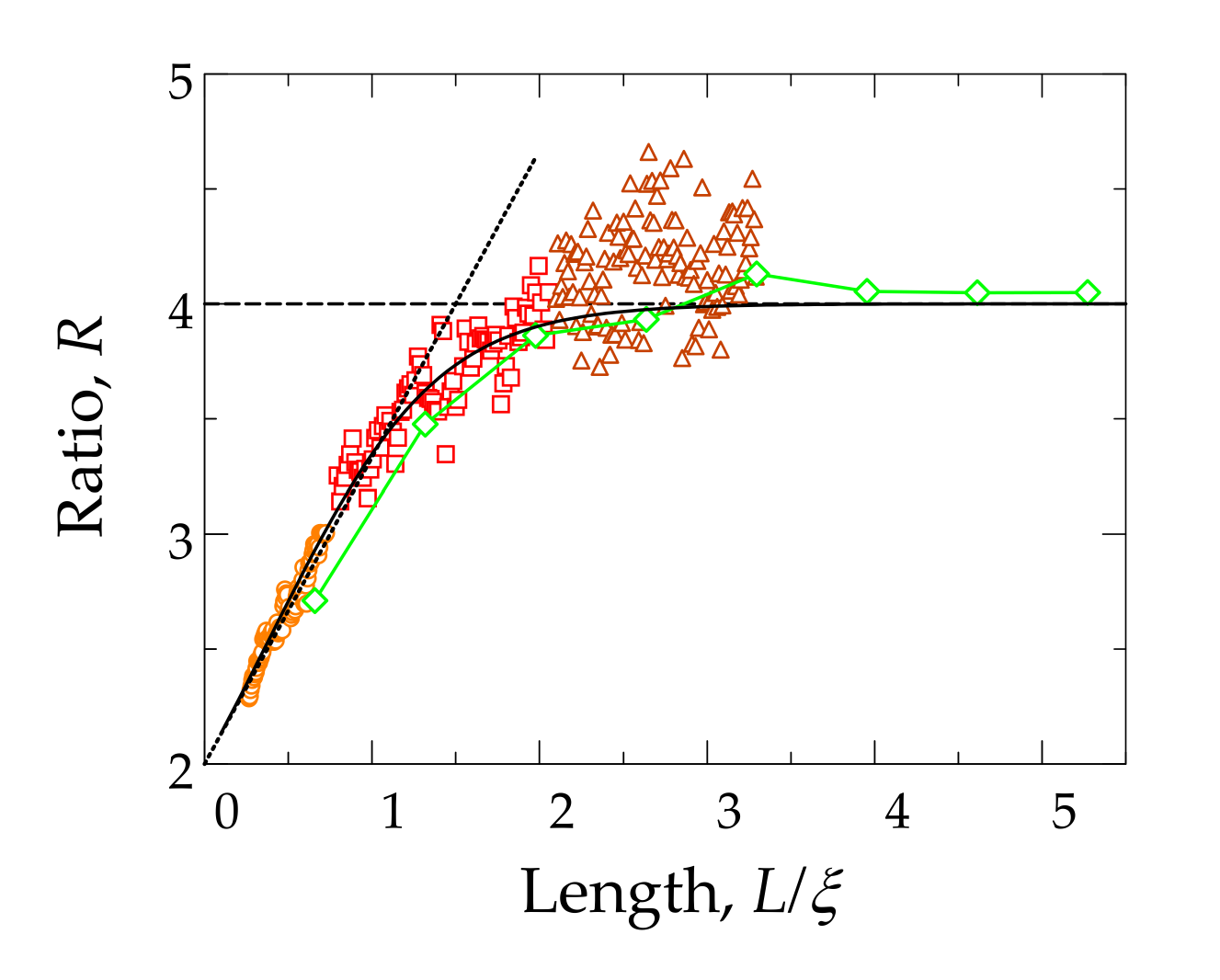}
\caption{ {\bf Crossover to the single-channel regime of transport.} The crossover to the single-channel regime is charted in terms of the ratio $R$ of the statistical moments of the normalized transmitted intensity $s_{ab}=T_{ab}/\langle T_{ab}\rangle$ and the normalized transmittance $s=T/\langle T\rangle$, $R=\langle s_{ab}^2\rangle/\langle s^2\rangle$, as a function of $L/\xi$. From the microwave experiment, $R$ was obtained in Sample $A$ (orange circles), $C$ (red squares), and $D$ (brown triangles). The data points of the same color and style correspond to different frequencies in samples of the same filling fraction and length. In the numerical simulations, planar disordered waveguides of eight different lengths were considered at a single scattering frequency (green line-connected diamonds). Both the experimental and numerical data agree well with exact nonperturbative calculations of $R$ for a quasi-1D geometry shown by the black solid line. The black dotted line represents the perturbative limit of $R$, for $L/\xi\ll1$, $R=2+4L/3\xi$, and the black dashed line represents the single-channel value $R=4$ in the deeply localized limit.}
\end{figure}

\begin{figure}
\includegraphics[angle=0, scale=1.0, width=0.7\textwidth]{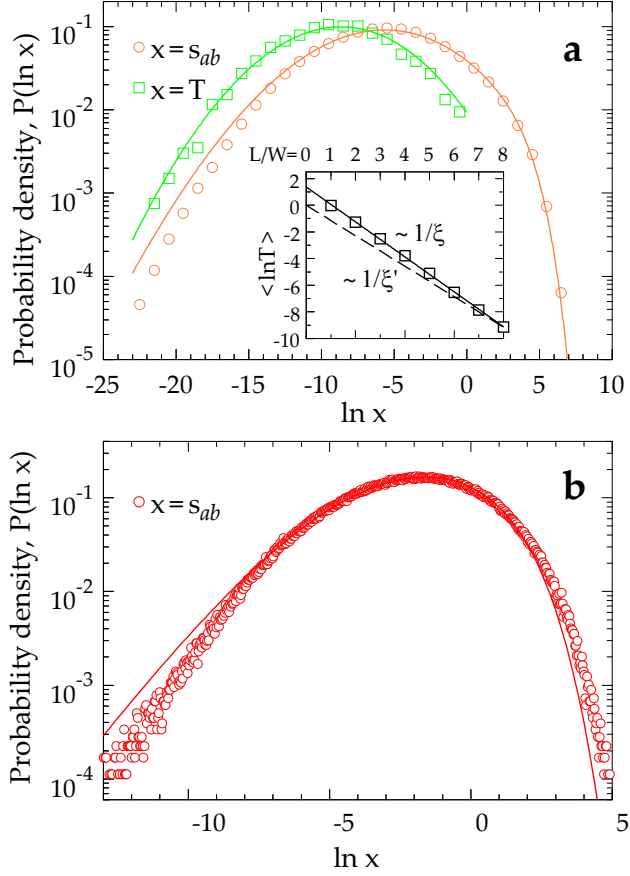}
\caption{
{\bf Statistics of the single-channel regime of transport.} {\bf a,} Probability density distributions $P(\ln T)$ (squares) and $P(\ln s_{ab})$ (circles) from the numerical data for a planar waveguide of $L/\xi=5.25$. The solid lines plotted through the data are the predictions from equations (2) and (4), respectively, with $L/\xi'=-\langle\ln T\rangle/2=4.57$. Inset: $\langle\ln T\rangle$ versus $L/W$ in the planar waveguides of eight different lengths (squares). The solid line is the best linear fit to the data, which yields the localization length $\xi=1.52W$. The broken line is $\langle\ln T\rangle =-2L/\xi'$ for the planar waveguide of $L=8W$, furnishing the renormalized localization length $\xi'=1.74W$. {\bf b,} Experimental results and prediction for $P(\ln s_{ab})$ in the quasi-1D system of $L/\xi=2.52$ (Sample $D$). Here, $L/\xi'=1.25$ is obtained from fitting the bulk of the measured distribution (circles) with $P(\ln s_{ab})$ from equation (4) (solid line).
}
\end{figure}

\begin{figure}
\includegraphics[angle=0, scale=1.0, width=0.7 \textwidth]{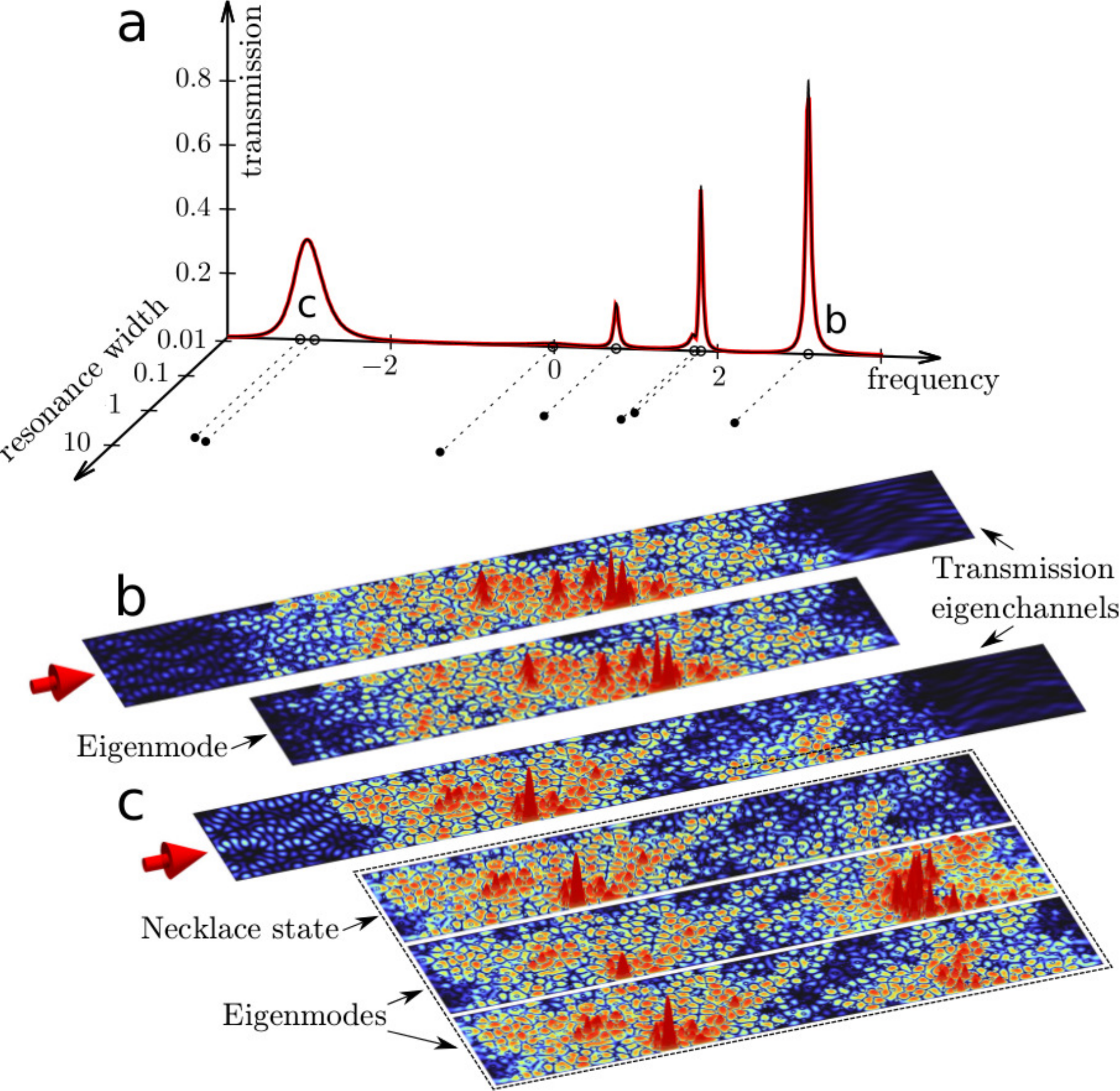}
\caption{
{\bf Transport through eigenchannels and eigenmodes in the single-channel regime.} {\bf a,} Numerically calculated transmittance $T$ (red line) versus detuning $\nu$ from the mid-frequency of the localization band, normalized by the average mode spacing $\Delta\nu$, in a disordered planar waveguide of $L/\xi=3.29$. The black line on top of the red curve shows the result of a fit of the transmittance, using the real [$\mbox{Re}\,\nu_m$] and imaginary [$\mbox{Im}\,\nu_m$] parts of the eigenfrequencies $\nu_m$ of the internal modes of the disordered region as fixed parameters (see empty and solid circles, respectively). Isolated eigenfrequencies represent individual localized modes ($b$), whereas closely spaced eigenfrequencies correspond to spectrally overlapping modes identified as necklace states ($c$). A spectral separation between the neighboring necklace states of about $5\Delta\nu$ can be noticed. {\bf b,} Spatial intensity pattern of the scattering state of the transmission eigenchannel (upper panel) and of the individual localized mode (lower panel) at the resonance peak $b$. {\bf c,} Spatial intensity profile of the scattering state of the transmission eigenchannel (upper panel) and of the two-mode necklace state (top panel in the framed box) at the resonance peak $c$. The lower two panels in the framed box display the two eigenmodes of the two-mode superposition.
}
\end{figure}

\begin{figure}
\subfigure{ \includegraphics[angle=0, scale=1.0, width=0.4\textwidth]{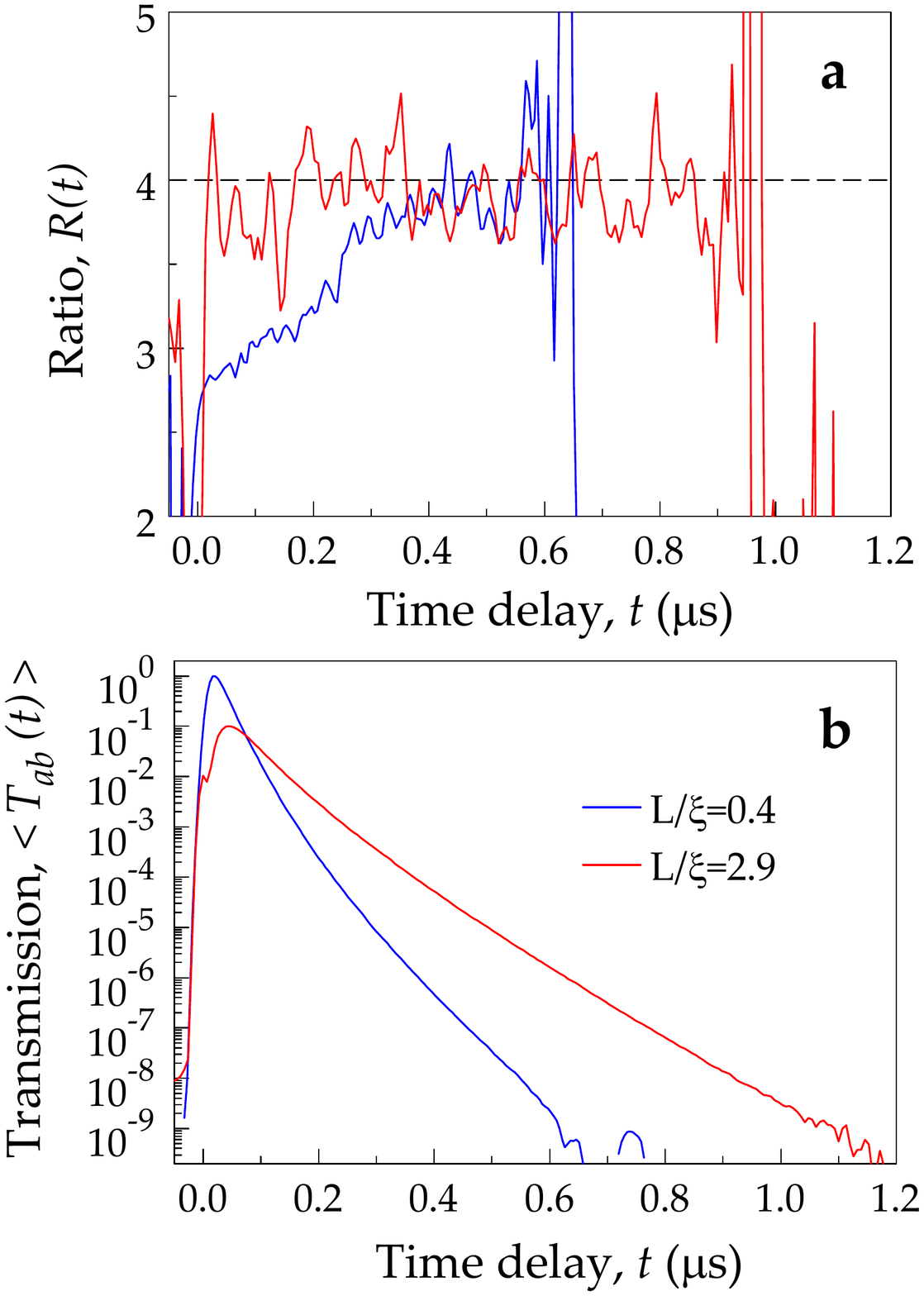}}
\subfigure{ \includegraphics[angle=0, scale=1.0, width=0.4\textwidth]{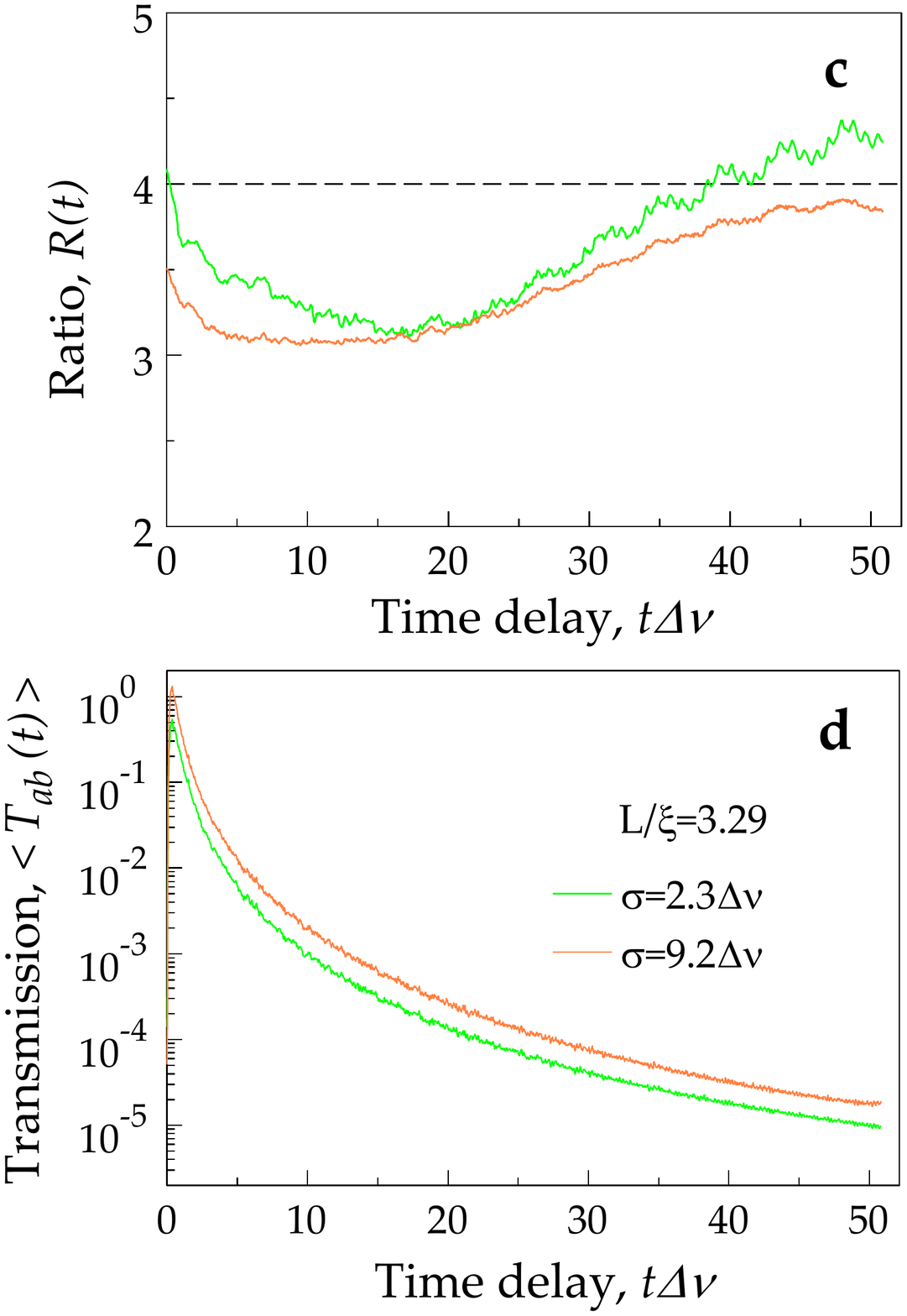}}
\caption{
{\bf Single-channel regime in the time domain.} Time-dependent ratio $R(t)$ in random quasi-1D ({\bf a,} experiment) and planar 2D ({\bf c,} numerics) waveguide systems following a Gaussian pulsed excitation of bandwidth $\sigma$. The horizontal dashed lines indicate $R=4$ of the single-channel regime. The experimental data in {\bf a} are for a localized sample of $L/\xi=2.9$ (Sample $D$, red line) and for a diffusive sample of $L/\xi=0.4$ (Sample $D$, blue line), using a pulse bandwidth of $\sigma=5\delta\nu=0.5\Delta\nu$ and $\sigma=1.8\delta\nu=3.9\Delta\nu$, respectively. Note that for the localized sample a single transmission eigenchannel dominates the pulsed transmission for all times, whereas a crossover to the single-channel regime with increasing time delay can be noticed in the diffusive system. The numerical data in {\bf c} are for a localized sample of $L/\xi=3.29$, using a bandwidth $\sigma=2.3\Delta\nu$ (green line) and $\sigma=9.2\Delta\nu$ (brown line). Note that for both cases we have $\sigma>\Delta\nu$, for which the single-channel regime sets in at long time delays. In addition, for the case where $\sigma $ is less than the average separation between neighboring necklace-state resonances of $5\Delta\nu$, the single-channel regime can be realized by transmission through a necklace state at short time delays (see the green line at $R=4$ for small $t$). For all cases, the average pulsed transmission, $\langle T_{ab}(t)\rangle$, is shown in {\bf b} (experiment) and {\bf d} (numerics). The experimental transmission curves were normalized to have a peak of unity and the curve for the localized system was displaced by a decade  for clarity of presentation.
}
\end{figure}


\begin{thebibliography}{99}
\bibitem{note} Note that the ratio $\langle s_{a}^2\rangle/\langle s^2\rangle$ could also be used to chart the crossover to the single-channel regime.
\end{thebibliography}
\end{document}